\documentclass[letterpaper,10pt]{article}   %% LaTeX 2e (preferred)
\usepackage{opex3} %% do not use with REVTeX4

\usepackage{here}
\usepackage{graphicx}

\usepackage{bm}% bold math

\begin{document}

\title{Coupled-resonator-induced reflection
in photonic-crystal waveguide structures}

\author{Sergei F. Mingaleev$^1$, Andrey E. Miroshnichenko$^{2}$, \\
and Yuri S. Kivshar$^2$}

\address{$^1$ VPI Development Center, Belarus High
Technologies Park, Minsk 220034, Belarus \\
$^2$Nonlinear Physics Center and Center for Ultra-high bandwidth
Devices for Optical Systems (CUDOS), Australian National University,
Canberra ACT 0200, Australia}

\begin{abstract}
We study the resonant transmission of light in a coupled-resonator
optical waveguide interacting with two nearly identical side
cavities. We reveal and describe a novel effect of the
coupled-resonator-induced reflection (CRIR) characterized by a very
high and easily tunable quality factor of the reflection line, for
the case of the inter-site coupling between the cavities and the
waveguide. This effect differs sharply from the
coupled-resonator-induced transparency (CRIT) -- an all-optical
analogue of the electromagnetically-induced transparency -- which
has recently been studied theoretically and observed experimentally
for the structures based on micro-ring resonators and photonic
crystal cavities. Both CRIR and CRIT effects have the same physical
origin which can be attributed to the Fano-Feshbach resonances in
the systems exhibiting more than one resonance. We discuss the
applicability of the novel CRIR effect to the control of the
slow-light propagation and low-threshold all-optical switching.
\end{abstract}


\section{Introduction}

\begin{figure}[t]
\centering\includegraphics[width=0.5\columnwidth]{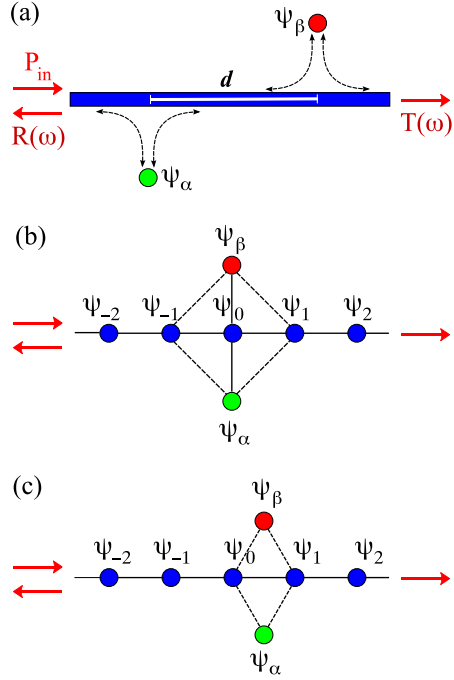}
\caption{Three types of the geometries of a straight
photonic-crystal waveguide side coupled to two nonlinear optical
resonators, $\alpha$ and $\beta$. Standard coupled-mode theory is
based on the geometry (a) which does not account for
discreteness-induced effects in the photonic-crystal waveguides. For
instance, light transmission and bistability are qualitatively
different for (b) on-site and (c) inter-site locations of the
resonator along waveguide and this cannot be distinguished within
the conceptual framework of structure of type (a).}
\label{fig:schematic}
\end{figure}

Many concepts of photonic devices employ high-$Q$ optical
resonators, such as micro-rings or photonic crystal cavities, side
coupled to the transmission waveguide. Among them, the devices which
employ \emph{several resonators} have attracted a special attention
because the coupling between optical resonators may lead to a
variety of novel effects such as the high-order resonances with
flattened passband region
\cite{Little:1997-998:JLT,Fan:1999-15882:PRB,Xu:2000-7389:PRE}, the
appearance of additional resonances with extremely high-$Q$ factors
\cite{Xu:2000-7389:PRE}, etc. One of the most interesting and
promising effects that
was discovered for the double-resonator photonic
structure shown schematically in Fig.~\ref{fig:schematic}(a)
is an all-optical analogue of the
\emph{electromagnetically-induced transparency} (EIT). We describe
the resonant transmission observed for such structure as the effect of
\emph{coupled-resonator-induced transparency} (CRIT)
\cite{Smith:2004-063804:PRA}. It has been predicted theoretically by
several research groups
\cite{Smith:2004-063804:PRA,Smith:2004-2503:JMO,Maleki:2004-626:OL,Matsko:2004-2515:JMO,Suh:2004-1511:IQE},
although one can also mention the early work
\cite{Opatrny:2001-23805:PRA} which suggested an idea of
\emph{macroscopic} double-resonator optical system exhibiting the
same EIT-like effect. Recently, the CRIT effect has been observed
experimentally in the system of two interacting micro-resonators for
the whispering-gallery modes~\cite{Naweed:2005-043804:PRA} and in
the integrated photonic chips employed either two micro-ring
resonators~\cite{Xu:2006-123901:PRL,Xu:2006-6463:OE} or two
photonic-crystal cavities~\cite{Pan:2007:PROC}. Such CRIT devices
provide an efficiently tunable `transparency on an optical chip',
and they are considered as a crucial step towards the development of
integrated all-optical chips~\cite{Boyd:2006-701:NAT}. In
particular, they can be employed for significant ($10^3$ times)
reduction of the threshold power for optical
bistability~\cite{Maes:2005-1778:JOSB}.

In this paper, we study the transmission of light in several types
of the resonant structures based on a photonic-crystal (PhC)
waveguide side-coupled to two nearly identical PhC cavities, as
shown schematically in Fig.~\ref{fig:schematic}(b) and
Fig.~\ref{fig:schematic}(c).
We confirm that in the case of \emph{on-site} coupling of two cavities
to a PhC waveguide, as shown schematically in Fig.~\ref{fig:schematic}(b),
the CRIT effect may be observed (see also the recent experimental
observation~\cite{Pan:2007:PROC} of the CRIT effect in a PhC structure
with on-site coupling).
However, for the structure shown in
Fig.~\ref{fig:schematic}(c), we reveal the existence  of a closely
related effect of \emph{coupled-resonator-induced reflection} (CRIR)
that manifests itself in an extremely narrow resonant reflection
line whose quality factor can easily be tuned by changing one of the
cavities. Specifically, the quality factor grows indefinitely when
the optical properties of two coupled cavities become identical.

In a sharp contrast to the CRIT effect that can be observed in the
structures with many different types of waveguides~
\cite{Xu:2006-123901:PRL,Xu:2006-6463:OE,Pan:2007:PROC} or even for
light propagating in free space with no waveguide at all
\cite{Naweed:2005-043804:PRA}, we reveal that the existence of the
CRIR effect is determined by the \emph{discrete nature} of the
photonic crystal waveguide and it can be observed, therefore, in
either coupled-resonator optical waveguides (CROWs) or PhC
waveguides, both created by an array of coupled optical cavities.
Recently, we have demonstrated a crucial importance of the discrete
nature of PhC waveguides for achieving high-$Q$ resonant reflection
and low-threshold all-optical switching in the slow-light
regime~\cite{Mingaleev:2006-046603:PRE,Mingaleev:2007-12380:OE}. We
have shown that this can be achieved by employing of
\emph{inter-site coupling} of a single Kerr nonlinear cavity to a
PhC waveguide.

%%-------------------------------------------------------------------
%%-------------------------------------------------------------------

Here, we extend those results further and demonstrate that
employing of \emph{inter-site coupling} of \emph{two cavities} in
the PhC waveguide structure shown schematically in
Fig.~\ref{fig:schematic}(c), leads to the transformation of the CRIT
effect into the CRIR effect. We also show that in
contrast to the CRIT effect in the structures with the on-site
coupling, the CRIR effect observed
in the structures with the inter-site coupling
\emph{survives in the slow-light regime}, and
thus it can be employed for a control on the slow-light propagation
and switching. As an example, we demonstrate a possibility to
achieve 100\% all-optical switching of slow light in the nonlinear
regime.

\section{Coupled-resonator-induced transparency}
\label{sec:CRIT}

First, we remind the basic facts about the CRIT effect assuming for
definiteness that the resonant photonic structure is created by a
straight dielectric waveguide and two side-coupled cavities, denoted
as $\alpha$ and $\beta$ and separated by the distance $d$ along waveguide, as shown
schematically in Fig.~\ref{fig:schematic}(a).

%%-------------------------------------------------------------------
\begin{figure}[t]
\centering
\includegraphics[width=0.95\columnwidth]{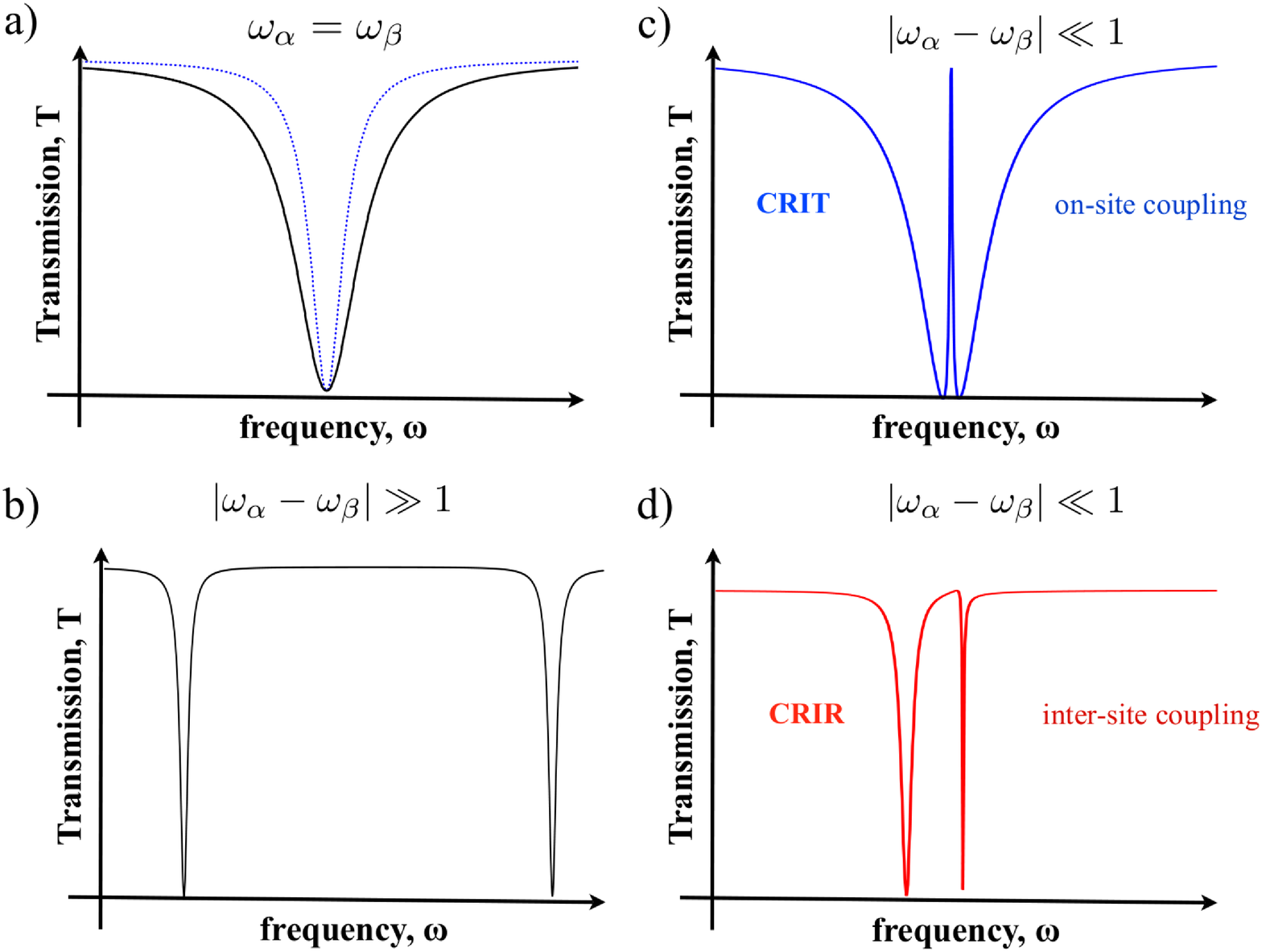}
\caption{Typical transmission curves for four different cases (a)
two identical side-coupled defects $\omega_\alpha=\omega_\beta$
(solid). For reference we also put transmission for single
side-coupled cavity (dashed); (b) two side-coupled cavities with
larger detuned eigenfrequencies $|\omega_\alpha-\omega_\beta|\gg1$;
(c) and (d) correspond to two side-coupled cavities with slightly
detuned eigenfrequencies $|\omega_\alpha-\omega_\beta|\ll1$ for two
geometries: (c) on-site coupling shown in Fig.~\ref{fig:schematic}(b)
and (d) inter-site coupling shown in Fig.~\ref{fig:schematic}(c).}
\label{fig:fig2-ok}
\end{figure}
%%-------------------------------------------------------------------

To simplify the analysis, below we assume that any losses are absent, so
that the transmission ($T(\omega)$) and reflection ($R(\omega)$)
coefficients can be conveniently written in the form
\begin{eqnarray}
\label{tr-Fano-abs} T(\omega)=
\frac{\sigma^{2}(\omega)}{\sigma^{2}(\omega) + 1} \quad \; , \quad
R(\omega)= \frac{1}{\sigma^{2}(\omega)+1} \; ,
\end{eqnarray}
where the detuning function $\sigma(\omega)$ may have a rather
general frequency dependence defined by the type of the PhC
waveguide-cavity structures (see, e.g., several examples in
Refs.~\cite{Mingaleev:2006-046603:PRE,Miroshnichenko:2005-036626:PRE}.
The zero transmission (complete reflection) corresponds to the
condition $\sigma(\omega)=0$, whereas the complete transmission
(vanishing reflection) corresponds to the condition
$\sigma(\omega)=\pm \infty$. Therefore, the resonant frequencies can
be conveniently found as zeros of the nominator and denominator of
the auxiliary function $\sigma(\omega)$.

For the simplest resonant structure created by a straight waveguide
coupled to a single cavity (e.g., the cavity $\alpha$),  we can
obtain~\cite{Mingaleev:2006-046603:PRE}
\begin{eqnarray}\label{sigma-one-cavity-strip}
\sigma(\omega) \simeq
\frac{(\omega_{\alpha}-\omega)}{\gamma_{\alpha}} \; ,
\end{eqnarray}
with a resonant reflection at the frequency that almost coincides
with the frequency $\omega_{\alpha}$ of the localized cavity mode of
an isolated cavity. The spectral width $\gamma_{\alpha}$ of the
resonance is determined by the overlap integral between the cavity
mode and the waveguide mode at the resonant frequency, rapidly
decaying as the distance between the cavity and the waveguide grows.

To find the characteristic function $\sigma(\omega)$ for the
waveguide structure with two cavities, we can employ a variety of
methods
\cite{Smith:2004-2503:JMO,Maleki:2004-626:OL,Matsko:2004-2515:JMO,Suh:2004-1511:IQE,Maes:2005-1778:JOSB},
including the simplest approach based on the transfer-matrix
technique~\cite{Fan:2002-908:APL}. A detailed analysis of the light
scattering in such structures can be found in
Refs.~\cite{Maleki:2004-626:OL,Xu:2006-123901:PRL,fan_physica}, and
here we present the results for the special case when two cavities
are separated by the distance $d = 2\pi m/k(\omega_{t})$, where
$k(\omega)$ is the waveguide dispersion relation, $m$ is integer,
and the frequency $\omega_{t}$ is defined below. In this case,
assuming that there is no direct coupling between the cavities, i.e.
either the distance $d$ is sufficiently large or the cavities are
coupled to the waveguide from the opposite sides, we obtain
\begin{eqnarray}\label{sigma-two-cavity-d0}
\sigma(\omega) \simeq
\frac{(\omega_{\alpha}-\omega)(\omega_{\beta}-\omega)}{\Gamma
(\omega_{t}-\omega)} \; ,
\end{eqnarray}
with the total resonance width $\Gamma = \gamma_{\alpha}+
\gamma_{\beta}$ and the frequency of perfect transmission
\begin{eqnarray}\label{params-two-cavity-d0}
\omega_t = \frac{\gamma_{\alpha}\omega_{\beta}+
\gamma_{\beta}\omega_{\alpha}}{\gamma_{\alpha} + \gamma_{\beta}} \;
,
\end{eqnarray}
lying in between the two frequencies, $\omega_{\alpha}$ and
$\omega_{\beta}$, of the zero transmission. Importantly, these
frequencies coincide with the frequencies of the localized cavity
modes of the isolated cavities $\alpha$ and $\beta$.

In the case when the cavities $\alpha$ and $\beta$ are identical,
Eq.~(\ref{sigma-two-cavity-d0}) reduces to the single-cavity
result~(\ref{sigma-one-cavity-strip}) with a doubled resonance
width, $\Gamma=2\gamma_{\alpha}$, as is illustrated in
Fig.~\ref{fig:fig2-ok}(a).
When the cavities $\alpha$ and $\beta$ are significantly different,
the structure exhibits two almost uncoupled single-cavity resonances,
as is illustrated in Fig.~\ref{fig:fig2-ok}(b).
However, introducing only a \emph{small}
difference between the parameters of the two cavities leads to opening an extremely
narrow resonant transmission on the background of the broader
reflection line, as is illustrated in Fig.~\ref{fig:fig2-ok}(c).
Indeed, for a small difference between the cavity parameters we
assume that $\gamma_{\beta} \approx \gamma_{\alpha}$, and
introducing notation $\delta\omega = \omega_{\beta} -
\omega_{\alpha}$, rewrite Eq.~(\ref{sigma-two-cavity-d0}) in the
vicinity of the resonant transmission frequency as
$\sigma(\omega) \approx \Gamma_t/(\omega-\omega_t)$
with $\omega_t=\omega_{\alpha}+\delta\omega/2$ and
\begin{eqnarray}
\label{gamma-narrow-CRIT}
\Gamma_t=\delta\omega^2/8\gamma_{\alpha} \; .
\end{eqnarray}
The line width $\Gamma_t$ of this resonance
can easily be controlled by tuning the frequency difference
$\delta\omega$, and the corresponding quality factor, $Q_t=\omega_t/2\Gamma_t
\approx 4\gamma_{\alpha}\omega_{\alpha}/\delta\omega^2$, grows
indefinitely when $\delta\omega$ vanishes. As we mentioned above,
this effect can be regarded as an all-optical analogue of the
electromagnetically-induced transparency, and it is now often
referred as the effect of \emph{coupled-resonator-induced
transparency} -- CRIT~\cite{Smith:2004-063804:PRA}.

\section{Model description of photonic crystal structures}
\label{sec:model}

For the photonic structures based on photonic-crystal waveguides we
have a new option (comparing with strip-waveguide structures)
for achieving the resonant transmission by placing
the cavity at different location relative to the waveguide and thus
exploring the \emph{discrete nature} of the structure. Recently, we
have demonstrated
\cite{Mingaleev:2006-046603:PRE,Mingaleev:2007-12380:OE} that, by
employing this extra degree of freedom in the waveguide-cavity
geometry with one cavity, we can dramatically increase the quality
factor of the resonant reflection in the slow-light regime and,
accordingly, decrease the light power required for the observation
of bistability in the light transmission. Here, we extend this
analysis to the case of two-cavity structures and reveal new
physics.

As was shown in
Refs.~\cite{Mingaleev:2006-046603:PRE,Miroshnichenko:2005-036626:PRE},
the light transmission in the waveguide-cavity photonic crystal
structures can be accurately modeled with the effective discrete
equations for the frequency-dependent dimensionless electric field
amplitudes of the cavity modes composing the waveguide,
$\psi_n(\omega)$, with integer $n \in [-\infty,\infty]$, and
side-coupled cavity modes, $\psi_{\mu}(\omega)$, with $\mu=\alpha,
\beta$. By applying this approach to the structures shown in
Figs.~\ref{fig:schematic}(b,c), such equations can be rewritten as
follows
\begin{eqnarray}
\label{eqs-motion}
\rho(\omega) \psi_n &=& \psi_{n+1} + \psi_{n-1}
+ \sum_{\mu=\alpha,\beta} \eta_{n,\mu} \psi_{\mu} \; , \nonumber \\
\frac{(\omega-\omega_{\mu})}{\gamma_{\mu}^{(0)}} \psi_{\mu} &=&
\sum_{n=-\infty}^{\infty} \eta_{n,\mu} \psi_{n} +
\lambda_{\mu} |\psi_{\mu}|^2 \psi_{\mu} \; .
\end{eqnarray}
Here, $\rho(\omega)=2\cos[k(\omega)s]$, where $k(\omega)$ describes
the waveguide dispersion and $s$ is the distance between
nearest-neighboring waveguide's cavities. Each side-coupled cavity
$\mu$ is described by the frequency $\omega_{\mu}$ of the localized
cavity mode, the dimensionless Kerr nonlinearity coefficient
$\lambda_{\mu}$, and the ``effective'' spectral width
$\gamma_{\mu}^{(0)}$ of the single-cavity resonance (the meaning of
``effective'' here will be clarified in the subsequent analysis).
The dimensionless coupling coefficients $\eta_{n,\mu}$ are assumed
to be equal (after appropriate rescaling of $\psi_{\mu}$,
$\gamma_{\mu}^{(0)}$, and $\lambda_{\mu}$) to either zero or $\pm
1$; their values will be indicated in what follows individually for
each studied case. The resonance detuning function $\sigma(\omega)$
can be found by solving Eq.~(\ref{eqs-motion}) in the way similar to
that described in
Refs.~\cite{Mingaleev:2006-046603:PRE,Miroshnichenko:2005-036626:PRE}.

These equations are valid in the approximation of local coupling
between the cavities composing the waveguide structure which has
been shown to produce qualitatively (and even semi-quantitatively)
correct results
\cite{Mingaleev:2006-046603:PRE,Miroshnichenko:2005-036626:PRE}. In
the form~(\ref{eqs-motion}), the effective discrete equations look
very similar to those derived from the coupled-mode theory
\cite{Yanik:2004-083901:PRL} for the case of coupled-resonator
optical waveguides~\cite{Yariv:1999-711:OL} and, therefore, our
subsequent analysis can be applied to all such structures as well.

First, we analyze briefly the photonic crystal structure with two
cavities, $\alpha$ and $\beta$, side-coupled to the same
\emph{on-site location} along PhC waveguide as shown in
Fig.~\ref{fig:schematic}(b). The situation when only one of these
cavities is coupled to the waveguide (with
$\eta_{n,\alpha}=\delta_{n,0}$ and $\eta_{n,\beta}=0$, where
$\delta_{n,m}$ is the Kronecker symbol) has already been studied
earlier in
Refs.~\cite{Mingaleev:2006-046603:PRE,Miroshnichenko:2005-036626:PRE}.
In this case the model parameters introduced in
Eq.~(\ref{eqs-motion}) are related to the parameters introduced in
Eqs.~(21)--(24) of Ref.~\cite{Mingaleev:2006-046603:PRE} as $\psi_n
\equiv A_n$, $\psi_{\alpha} \equiv (V_{0,\alpha}/V_{1w})
A_{\alpha}$, $\gamma_{\alpha}^{(0)}=\nu_{\alpha} \omega_{\alpha}
V_{0,\alpha} V_{\alpha,0} / V_{1w}$, and $\lambda_{\alpha} =
(\kappa_{\alpha}
\chi_{\alpha}^{(3)}/V_{\alpha,0})(V_{1w}/V_{0,\alpha})^3$.
Therefore, the linear (at $\lambda_{\alpha}=0$) detuning function
$\sigma(\omega)$ found for this case in
Ref.~\cite{Mingaleev:2006-046603:PRE} takes the form of
Eq.~(\ref{sigma-one-cavity-strip}) with unchanged form
of $\omega_{\alpha}$ and with
$\gamma_{\alpha}=\gamma_{\alpha}^{(0)}/\sin[k(\omega_{\alpha})s]$.
As one can see, $\gamma_{\alpha}^{(0)}$ is the spectral width of the
resonant reflection line that would be produced by a single on-site
side-coupled cavity in the case when its frequency $\omega_{\alpha}$
lies at the center of the waveguide transmission band,
$k(\omega_{\alpha})=\pi/2s$.

In the same way, we can show that the detuning function of the
on-site two-cavity structure with $\eta_{n,\mu}=\delta_{n,0}$ for
both $\mu$, shown in Fig.~\ref{fig:schematic}(b), takes the form of
Eqs.~(\ref{sigma-two-cavity-d0})--(\ref{gamma-narrow-CRIT}) with
unchanged forms of $\omega_{\alpha}$ and $\omega_{\beta}$, and with
$\gamma_{\alpha}=\gamma_{\alpha}^{(0)}/\sin[k(\omega_{\alpha})s]$
and $\gamma_{\beta}=\gamma_{\beta}^{(0)}/\sin[k(\omega_{\beta})s]$.
Correspondingly, this \emph{on-site two-cavity structure exhibits
the same effect of coupled-resonator-induced transparency},
illustrated in Fig.~\ref{fig:fig2-ok}(c), as discussed in the
previous section.

\section{Coupled-resonator-induced reflection}
\label{sec:CRIR}

Now let us analyze an alternative photonic crystal structure
with two cavities, $\alpha$ and $\beta$, side-coupled
to the same \emph{inter-site location} along PhC waveguide as
shown in Fig.~\ref{fig:schematic}(c).
In this case, $\eta_{n,\mu}=(\delta_{n,0} \pm \delta_{n,1})$,
where the upper sign in ``$\pm$'' corresponds to the even-symmetry
cavity modes, while the bottom sign corresponds to the odd-symmetry
cavity modes.

The situation when only one of the cavities (say, the cavity
$\alpha$) is coupled to the waveguide (with
$\eta_{n,\alpha}=(\delta_{n,0} \pm \delta_{n,1})$, but
$\eta_{n,\beta}=0$) has already been studied for the even-symmetry
cavity modes in
Refs.~\cite{Mingaleev:2006-046603:PRE,Miroshnichenko:2005-036626:PRE}.
Extending those results to the case of both even- and odd-symmetry
cavity modes, the detuning parameter for the inter-site
single-cavity structure can be obtained in the form
\begin{equation}
\label{sigma-one-cavity} \sigma(\omega) \simeq
\frac{(\tilde{\omega}_{\alpha}^{\pm}-\omega)}{\gamma_{\alpha}^{\pm}} \; ,
\end{equation}
where the resonant reflection frequency
$\tilde{\omega}_{\alpha}^{\pm} = \omega_{\alpha} \mp
\gamma_{\alpha}^{(0)}$ is shifted to one or the other side from the
frequency $\omega_{\alpha}$ of the cavity mode, depending on its
symmetry. The spectral width $\gamma_{\alpha}^{\pm} =
\gamma_{\alpha}^{(0)} / \left[
\tan(k(\tilde{\omega}_{\alpha}^{\pm})s/2)\right]^{\pm 1}$ of this
resonance line equals to $\gamma_{\alpha}^{(0)}$ at the center of
waveguide pass-band, but it vanishes at one of the band edges,
giving birth to \emph{extremely high-quality resonant reflection
lines in the slow-light regime}
\cite{Mingaleev:2006-046603:PRE,Mingaleev:2007-12380:OE}.

%%-------------------------------------------------------------------
\begin{figure}[t]
\centering
\includegraphics[width=0.96\columnwidth]{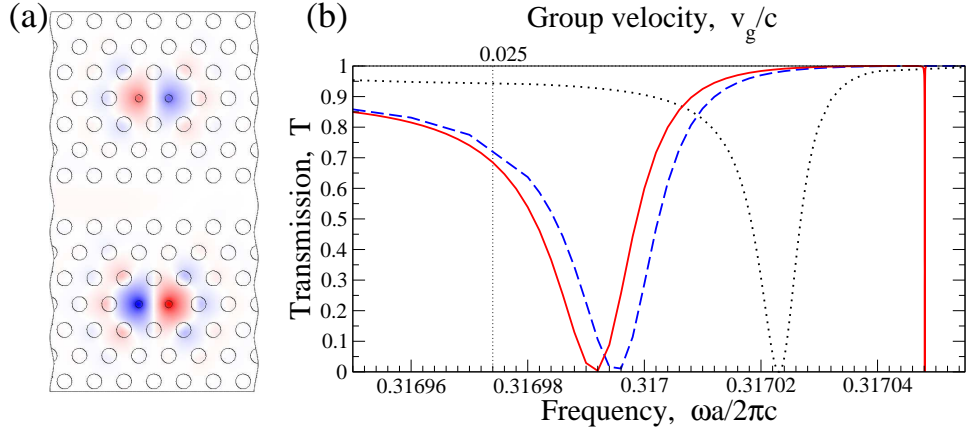}
\caption{(a) Schematic structure of a photonic-crystal waveguide
coupled to two side cavities. (b) Linear transmission coefficient of
the photonic-crystal waveguide for: (i) single side-couple cavity
with $\epsilon_{\alpha}=12.$ (dotted); (ii) two identical
side-couple cavities with $\epsilon_{\alpha}=\epsilon_{\beta}=12.$
(dashed); (iii) two side-coupled cavities with different
permittivities $\epsilon_{\alpha}=12.$ and $\epsilon_{\beta}=12.001$
(solid).} \label{fig:fig2}
\end{figure}
%%-------------------------------------------------------------------

For the problem of the light transmission in the inter-site
two-cavity structure shown in Fig.~\ref{fig:schematic}(c), we solve
Eq.~(\ref{eqs-motion}) and obtain the detuning parameter in the
form,
\begin{eqnarray}\label{sigma-two-cavity-CRIR}
\sigma(\omega) \simeq
\frac{(\omega_{r1}^{\pm}-\omega)(\omega_{r2}^{\pm}-\omega)}{\Gamma^{\pm}
(\omega_{t}-\omega)} \; ,
\end{eqnarray}
that looks qualitatively similar to Eq.~(\ref{sigma-two-cavity-d0}).
Moreover, the frequency $\omega_t$ of the perfect transmission is
determined by the same  equation for the CRIT effect
~(\ref{params-two-cavity-d0}) with $\gamma_{\mu} =
\gamma_{\mu}^{(0)}$. However, now the total resonance width
$\Gamma^{\pm} = \left[ \tan(k(\omega_{t}) s/2) \right]^{\mp 1}
(\gamma_{\alpha}^{(0)}+ \gamma_{\beta}^{(0)})$ depends on the
waveguide dispersion, $k(\omega_{t})$, at the resonance frequency
leading to the same enhancement of the resonance quality factor at
one of the edges of waveguide pass-band as obtained above for the
inter-site one-cavity structure. More importantly, the resonant
reflection frequencies,
\begin{eqnarray}
\label{omega-r-two-cavity} \omega_{r1}^{\pm} &=&
\frac{\omega_{\alpha}+\omega_{\beta}}{2} \mp
\frac{\gamma_{\alpha}^{(0)}+\gamma_{\beta}^{(0)}}{2} \nonumber \\ &+&
\frac{1}{2} \sqrt{(\omega_{\alpha}-\omega_{\beta} \mp
\gamma_{\alpha}^{(0)} \pm \gamma_{\beta}^{(0)})^2 + 4 \gamma_{\alpha}^{(0)}
\gamma_{\beta}^{(0)}}
\; , \nonumber \\
\omega_{r2}^{\pm} &=& \frac{\omega_{\alpha}+\omega_{\beta}}{2} \mp
\frac{\gamma_{\alpha}^{(0)}+\gamma_{\beta}^{(0)}}{2} \nonumber \\ &-&
\frac{1}{2} \sqrt{(\omega_{\alpha}-\omega_{\beta} \mp
\gamma_{\alpha}^{(0)} \pm \gamma_{\beta}^{(0)})^2 + 4 \gamma_{\alpha}^{(0)}
\gamma_{\beta}^{(0)}} \; ,
\end{eqnarray}
do not coincide with the cavity-mode frequencies $\omega_{\alpha}$
and $\omega_{\beta}$. Moreover, these two resonant reflection
frequencies are always separated by a finite distance exceeding the
value $2\sqrt{\gamma_{\alpha}^{(0)} \gamma_{\beta}^{(0)}}$ and,
therefore, the existence of a narrow resonant transmission becomes
merely impossible.

%%%%%%%%%%%%%%%%%%%%%%%%%%%%%%%%%%%%%%%%%%

In contrast, the inter-coupling between the waveguide and two
cavities in this system manifests itself in {\em a qualitatively new
effect} of coupled-resonator-induced reflection: for small
$\delta\omega = \omega_{\beta} - \omega_{\alpha}$ one of the
resonant reflection frequencies moves very close to the perfect
transmission frequency, $\omega_{t}$, producing a \emph{narrow
resonant reflection line}, as is illustrated in
Fig.~\ref{fig:fig2-ok}(d). The frequency of this line is always
close to the frequency $\omega_{\alpha}$ of the cavity mode, while
its spectral width is determined by the frequency difference
$\delta\omega$ growing indefinitely as $\delta\omega$ vanishes. For
a small difference between the cavity frequencies,
assuming $\gamma_{\beta}^{(0)} \approx \gamma_{\alpha}^{(0)}$, we can
estimate that the spectral width of this narrow reflection line is
\begin{eqnarray}
\label{gamma-narrow-CRIR}
\Gamma_{r}^{\pm} &=& \left[ \tan(k(\omega_{t}) s/2) \right]^{\mp 1}
\gamma_{\alpha}^{(0)}\left( 1 -
1/\sqrt{1+(\delta\omega/2\gamma_{\alpha}^{(0)})^2} \right) \nonumber \\
&\approx&
\left[ \tan(k(\omega_{t}) s/2) \right]^{\mp 1}
\delta\omega^2/8\gamma_{\alpha}^{(0)} \; ,
\end{eqnarray}
At small $\delta\omega$ and $\omega_{t}$ lying at the center of the
passing band, this spectral width almost coincides with the
corresponding width~(\ref{gamma-narrow-CRIT}) of the narrow resonant
transmission line in the structures exhibiting the CRIT effect.

In addition to this narrow resonant reflection line, there always
exists the second resonant reflection line located (at small
$\delta\omega$) at the frequency $\omega_{r}^{\pm}=\omega_{\alpha}
\mp 2 \gamma_{\alpha}^{(0)}$, significantly shifted from the
frequency of the cavity modes. This line is characterized by the
spectral width $\Gamma^{\pm} = 2\gamma_{\alpha}^{(0)}\left[
\tan(k(\omega_{t}) s/2) \right]^{\mp 1}$ which is twice larger than
the width of the corresponding single-cavity resonance
\cite{Mingaleev:2006-046603:PRE}.

It should be emphasized that despite such a qualitative difference
in their spectral manifestations, both CRIT and CRIR effects have
the same physical origin which can be attributed to the
Fano-Feshbach resonances
\cite{Feshbach:1958-357:AP,Mies:1968-164:PR} known to originate from
the interaction of two or more resonances (e.g., two Fano
resonances) in the overlapping regime where the spectral widths of
the resonances are comparable to or larger than the frequency
separation between them. In a general case, this leads to a drastic
deformation of the transmission spectrum and the formation of
additional resonances with sharp peaks. The Fano-Feshbach resonances
are associated with a collective response of multiple interacting
resonant degrees of freedom, and they have numerous evidences in
quantum mechanical systems~\cite{Rotter:2003:PRB,Mies:2004:PRA}.

For the PhC structures studied in this paper, the two resonant
degrees of freedom are associated with two side-coupled cavities,
which can be coupled to PhC waveguide in two different ways,
illustrated in Fig.~\ref{fig:schematic}(b) and
Fig.~\ref{fig:schematic}(c), with the different manifestation of the
Fano-Feshbach resonance in these two cases in the form of either
CRIT or CRIR effects.

%%-------------------------------------------------------------------
\begin{figure}[t]
\centering
\includegraphics[width=1.0\columnwidth]{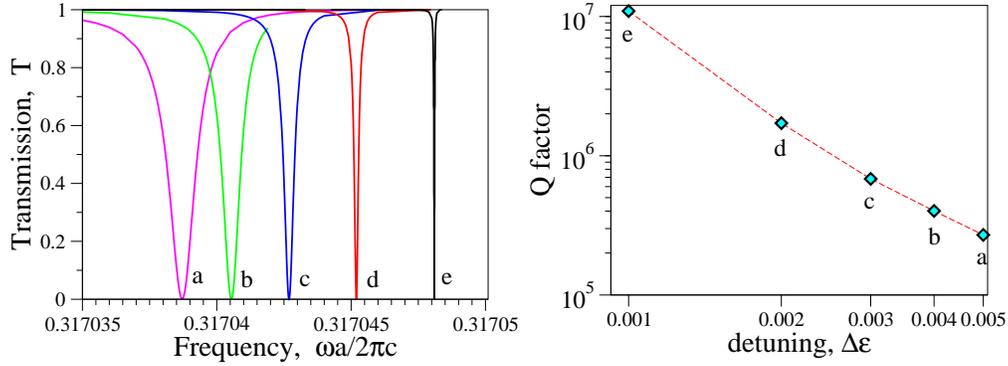}
\caption{Width of the asymmetric Fano-Feshbach resonance vs. the
permittivity detuning of two cavities.
In all cases $\epsilon_{\alpha}=12.$, and
(a) $\epsilon_{\beta}=12.005$, (b) $\epsilon_{\beta}=12.004$,
(c) $\epsilon_{\beta}=12.003$, (d) $\epsilon_{\beta}=12.002$,
(e) $\epsilon_{\beta}=12.001$,.
} \label{fig:fig3}
\end{figure}
%%-------------------------------------------------------------------

\section{Photonic structure exhibiting the CRIR effect}

As an example of the specific structure exhibiting the CRIR effect,
we consider a two-dimensional PhC composed of a triangular lattice
of dielectric rods in air. The rods are made of either Si or GaAs
($\varepsilon=12$) with the radius $r=0.25a$, where $a$ is the
lattice spacing. This type of PhC has two large bandgaps for the
E-polarized light (i.e. when the electric field is parallel to the
rods), and we employ the first gap between the frequencies $\omega
a/2 \pi c = 0.2440$ and $\omega a/2 \pi c=0.3705$.

We assume that the photonic-crystal waveguide is created by removing
a row of rods, while each of the cavities is created by reducing
radius of two nearest rods to the value $r_{\rm def}$.
The lines connecting defect rods in each cavity are
parallel to the waveguide being separated from the waveguide by
three rows of rods, as shown in Fig.~\ref{fig:fig2}(a). Each cavity
supports two localized modes: (i) high-frequency mode with an odd
symmetry, and (ii) low-frequency mode with an even symmetry. Here,
we use the properties of the odd-symmetry mode which has been
discussed recently for the slow-light
applications~\cite{Mingaleev:2007-12380:OE}, and, therefore, we
select the bottom sign in expressions ``$\pm$'' and ``$\mp$'' in
Eqs.~(\ref{sigma-one-cavity})--(\ref{gamma-narrow-CRIR}).

First, in Fig.~\ref{fig:fig2} and Fig.~\ref{fig:fig3} we present
numerically accurate results for this structure obtained by
employing the Wannier functions approach with eleven maximally
localized Wannier functions, in the way outlined in
Ref.~\cite{Busch:2003-R1233:JPCM}. To emphasize that the CRIR effect
in such a structure survives also in the slow-light regime, we shift
the resonance to a vicinity of the passing band edge $k=0$ by
choosing $r_{\rm def}=0.1213a$. In Fig.~\ref{fig:fig2}(a) we plot
the corresponding distributions of the electric field at the
low-frequency resonance. Figure~\ref{fig:fig2}(b) shows the
transmission spectra for three structures: with only a single
side-coupled cavity (dotted line), with two identical side-coupled
cavities (dashed line), and with two slightly different side-coupled
cavities (solid line). As can be seen, the results of these
full-scale calculations are in a very good agreement with our
analysis  based on the approximate discrete model and presented in
Sec.~\ref{sec:CRIR}.

In Fig.~\ref{fig:fig3} we plot the transmission spectra of the
narrow resonant reflection line and the corresponding resonance quality
factors for several values of detuning between the dielectric
constants of two cavities. As is seen, the indefinite growing of the
resonance quality factor with vanishing of this detuning is in full
qualitative agreement with the model prediction provided by
Eq.~(\ref{gamma-narrow-CRIR}).

\section{Nonlinear transmission and all-optical switching}

\begin{figure}[t]
\vspace{20pt} \centering
\includegraphics[width=0.65\columnwidth]{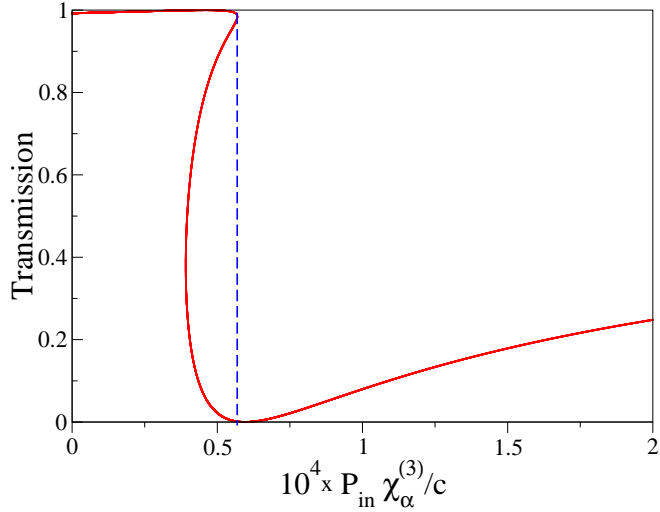}
\caption{Nonlinear transmission for the parameters used in
Fig.~\ref{fig:fig2}. We assume that one  cavity is nonlinear, such
that $\lambda_{\alpha} =  3.903 \cdot 10^7 \chi_{\alpha}^{(3)}$ and
$\lambda_{\beta}=0$. The light frequency is $\omega=0.3170418 (2\pi c/a)$.}
\label{fig:fig4}
\end{figure}

In the {\em nonlinear} regime, the CRIR structures described here
demonstrate low-threshold bistable transmission of light due to the
ultra-high $Q$ factor of the asymmetric Fano-Feshbach resonance. For
specific example, in Fig.~\ref{fig:fig4} we present the results for
the nonlinear transmission of the waveguide-two-cavities structure
for the parameters used in Fig.~\ref{fig:fig2} assuming that only
one of the cavities is made nonlinear. In this case, choosing the
frequency $\omega=0.3170418 (2\pi c/a)$ close to the asymmetric resonance
allows achieving a complete 100$\%$ switching in the regime of the
slow-light propagation.

%%-------------------------------------------------------------------
%%-------------------------------------------------------------------

We would like to mention that, to demonstrate the efficiency of the
approximations employed in this work, the results for the nonlinear
transmission shown in Fig.~\ref{fig:fig4} are obtained by employing
the approximate discrete model based on Eq.~(\ref{eqs-motion}) with
the model parameters calculated with the the Wannier function
approach~\cite{Busch:2003-R1233:JPCM}, in contrast to the numerical
results presented in Fig.~\ref{fig:fig2}--Fig.~\ref{fig:fig3}. Using
eleven maximally localized Wannier functions, we obtain the
following expansions,
\begin{equation}
\omega_{\alpha,\beta} \approx [ 0.317045 + 6.83 \cdot 10^{-3}
(\varepsilon_{\alpha,\beta}-12) ] (2\pi c/a),
\end{equation}
and
\begin{equation}
\gamma_{\alpha,\beta}^{(0)} \approx \pm\ (4.3 \cdot 10^{-6}) (2\pi
c/a),
\end{equation}
while the waveguide dispersion in the slow-light regime can be
approximated as
\begin{equation} \omega(k)\approx [0.31680 + 0.89626
(ka/2\pi)^2] (2\pi c/a).
\end{equation}

Thus, one of our major finding is that, in a sharp contrast to the
CRIT effect, this novel CRIR resonant effect survives also in the
slow-light regime and, due to its asymmetric line shape, it allows
achieving a 100\% all-optical switching of slow light in the
nonlinear regime. Currently, the slow-light applications of optical
structures based on photonic crystals attract a rapidly growing
attention due to the recently achieved experimental success in the
observation of slow-light propagation
\cite{vlasov,kuipers,Notomi:2001-253902:PRL,Jacobsen:2005-7861:OE}.
However, many of the device concepts suggested so far in the physics
of photonic crystals cannot be extended to the slow-light regime,
and it is important to reveal and analyze novel types of the design
concepts which would allow to perform useful operations such as
all-optical switching and routing with low group velocities
\cite{Assefa:2006-745:OL,Vlasov:2006-50:OL,Mingaleev:2006-046603:PRE,Mingaleev:2007-12380:OE}.
We therefore believe that the CRIR effect described in this work may
be especially useful for elaborated control of slow light, however
more extensive studies are beyond the scope of this paper.

\section{Conclusions}

We have analyzed the resonant transmission of light in a
coupled-resonator optical waveguide interacting with two nearly
identical cavities, paying a particular attention to differences
between the {\em on-site} and {\em inter-site} locations of the
cavities relative to the waveguide. When two cavities are strongly
detuned and associated resonances are well separated [see
Fig.~\ref{fig:fig2-ok}(b)], both types of photonic-crystal
structures are characterized by the similar transmission curves with
two distinct Fano resonances. However, when two cavities are only
slightly detuned and, therefore, the associated resonances strongly
overlap, the on-site and inter-site structures produce
quantitatively different transmission curves. Specifically, the
on-site geometry shown in Fig.~\ref{fig:schematic}(b) may lead to
the CRIT effect with a sharp symmetric resonant transmission line
[see Fig.~\ref{fig:fig2-ok}(c)], while the inter-site geometry shown
in Fig.~\ref{fig:schematic}(c) may lead to the CRIR effect with a
sharp and asymmetric resonant reflection line [see
Fig.~\ref{fig:fig2-ok}(d)]. The new CRIR transmission we have
described here is characterized by a very high and easily tunable
quality factor of the reflection line, and this effect differs
sharply from the CRIT effect being an all-optical analogue of the
electromagnetically-induced transparency demonstrated recently for
the structures based on micro-ring resonators. We have demonstrated
that the CRIR effect survives also in the slow-light regime and, due
to its asymmetric line shape, it allows achieving a 100\%
all-optical switching of slow light in the nonlinear regime.

\section*{Acknowledgements}

This work has been supported by the Australian Research Council
through the Discovery and Center of Excellence research projects.

\end{document}